\documentclass[12pt,draftclsnofoot, onecolumn]{IEEEtran}
\usepackage{graphicx,cite,epsfig,amssymb,amsmath}
\usepackage{color,xcolor}
\usepackage{pifont}
\usepackage{stmaryrd}
\usepackage{bbding}
\usepackage{subfigure}
\usepackage{setspace}
\usepackage{color}

\begin{document}
\begin{spacing}{1.6}

\markboth{IEEE Communications Magazine, Feature Topic on Wireless Physical Layer Security, June 2015.} {}

\title{Multi-Antenna Relay Aided Wireless Physical Layer Security}
\author{Xiaoming~Chen,~\IEEEmembership{Senior Member, IEEE}, Caijun~Zhong,~\IEEEmembership{Senior Member, IEEE}, Chau~Yuen,~\IEEEmembership{Senior Member, IEEE}, and Hsiao-Hwa~Chen,~\IEEEmembership{Fellow, IEEE}
\thanks{Xiaoming~Chen (e-mail: {\tt chenxiaoming@nuaa.edu.cn}) is
with the College of Electronic and Information Engineering, Nanjing
University of Aeronautics and Astronautics, China.
Caijun~Zhong (e-mail: {\tt caijunzhong@zju.edu.cn}) is with the
Department of Information Science and Electronic Engineering,
Zhejiang University, China. Chau~Yuen (e-mail: {\tt
yuenchau@sutd.edu.sg}) is with Singapore University of
Technology and Design, Singapore. Hsiao-Hwa Chen (email: {\tt
hshwchen@mail.ncku.edu.tw}) is with the Department of Engineering
Science, National Cheng Kung University, Taiwan.}
\date{\today}}
\renewcommand{\baselinestretch}{1.5}
\thispagestyle{empty} \maketitle

\begin{abstract}
With growing popularity of mobile Internet, providing secure wireless services has become a critical issue. Physical layer security (PHY-security) has been recognized as an effective means to enhance wireless security by exploiting wireless medium characteristics, e.g., fading, noise, and interference. A particularly interesting PHY-security technology is cooperative relay due to the fact that it helps to provide distributed diversity and shorten access distance. This article offers a tutorial on various multi-antenna relaying technologies to improve security at physical layer. The state of the art research results on multi-antenna relay aided PHY-security as well as some secrecy performance optimization schemes are presented. In particular, we focus on large-scale MIMO (LS-MIMO) relaying technology, which is effective to tackle various challenging issues for implementing wireless PHY-security, such as short-distance interception without eavesdropper channel state information (CSI) and with imperfect legitimate CSI. Moreover, the future directions are identified for further enhancement of secrecy performance.
\end{abstract}


\IEEEpeerreviewmaketitle

\section{Introduction}
We have witnessed significant growth in wireless communications
due to the rapid technological advancements in cellular, sensor,
cyber-physical, and machine-to-machine (M2M) communication networks.
Applications based on these diverse wireless systems are used to transmit and receive confidential/private data (e.g., credit card information, energy pricing, e-helath data, command and
control messages, etc.). Therefore, it is important to guarantee
secure communications in the presence of possible undesired third
parties, e.g., attackers, eavesdroppers, adversaries with malicious
data injection capability, etc. Traditionally, secure
communication systems were implemented using upper layer protocols
and tools, such as cryptography. However, cryptography requires an
extra secure channel for exchange of private keys. Note that
for mobile or unstructured networks, it is difficult to provide a
reliably secure channel. Recently, a new paradigm known as
PHY-security that exploits the randomness of wireless propagation
medium has emerged \cite{PLS1}. The benefits of PHY-security
are two-fold. First, heavy dependence on complex
higher-layer encryption may not be necessary, leaving more
computation resources for communications. Second, PHY-security
avoids the use of private keys, and thus it can be made more
applicable. And in practical systems, PHY-security can serve as an
additional layer of protection on top of the existing security
features.

From an information-theoretic viewpoint, the essence of
PHY-security is to maximize the performance difference between
legitimate and eavesdropper channels \cite{SC}. Generally
speaking, it aims to enhance the legitimate signal and impair the
eavesdropper signal simultaneously, thus realizing secure,
reliable, and QoS-guaranteed communications. In this context, a
variety of physical layer techniques can be utilized to enhance
wireless security. Wherein, multi-antenna technique is one of the
most powerful ways for secure communications. Making use of
spatial degrees of freedom, it is possible for us to increase the
legitimate channel rate and concurrently decrease the eavesdropper
channel rate. As a simple example, if signal is transmitted in the
null space of the eavesdropper channel, the eavesdropper cannot
receive any information, and thus information leakage is avoided.
It is worth pointing out that the quality of both legitimate and
interception signals are related largely to the propagation
distance. If the interception distance is short, it is difficult
to provide a high QoS-guaranteed secure communication even
exploiting the benefit of multi-antenna technique. This is because
the gain from multi-antenna technique is small compared to path
loss of signal propagation. To address this challenge, the relaying
technology was introduced into PHY-security, so as to shorten the
propagation distance of the legitimate signal \cite{Relay1}.
Especially, the multi-antenna relaying technology has attracted
considerable attention as it has the advantages of both
multi-antenna and relaying technologies.

To fully exploit the benefits of multi-antenna relaying technology
for PHY-security, it is necessary to adaptively adjust the
transmit parameters, such as transmit beams, transmit powers,
transmit durations, and relaying protocols \cite{Relay2}.
Intuitively, in order to implement these secrecy performance
optimization schemes, the transmitters require full or at least
partial CSI. Unlike traditional relaying systems, the secrecy
relaying system involves different types of CSI. In addition to the
legitimate CSI about source-relay and relay-destination channels,
there is the eavesdropper CSI about source-eavesdropper and
relay-eavesdropper channels. As revealed in the literatures, the
CSI has a great impact on the performance of adaptive transmission
techniques. If full CSI is available at the source and the relay,
it is possible to attain a steady secrecy rate, or even achieve the
secrecy capacity. However, eavesdropper CSI is usually
unavailable, since an eavesdropper can be passive and keeps silent. In
this case, it is impossible to provide a steady secrecy rate over
all realizations of fading channels. To this end, some new
performance metrics, i.e., ergodic secrecy rate, secrecy outage
probability, and interception probability, are proposed
accordingly to evaluate wireless security in a statistical sense
\cite{ImperfectCSI}. Moreover, legitimate CSI may also be imperfect as
normally it is obtained through limited feedback or by making use
of channel reciprocity. Under this condition, it is nontrivial to
design adaptive performance optimization schemes.

In this article, we intend to provide an overview on various
state-of-the-art multi-antenna relaying technologies from the
perspective of PHY-security. Especially, we investigate viable
secrecy performance optimization schemes in the framework of
multi-antenna secure relaying system. Then, we discuss and analyze
an up-to-date multi-antenna relaying technology,
namely large-scale MIMO (LS-MIMO) relaying, to show the benefits
of cooperative schemes for wireless security. At the end, we
conclude the whole article with a discussion on future research
directions on secure relaying systems.

\section{State-of-the-Art Multi-Antenna Secure Relaying Technologies}
\vspace{0.1in}

Some pioneering works on multi-antenna relaying technologies for
PHY-security revealed the fundamental functions on wireless
security. Specifically, the multi-antenna relay plays two roles:
\begin{enumerate}
\item To help the source by enhancing channel quality to the
legitimate destination;
\item To repress the interception by deteriorating the channel
condition to the eavesdropper.
\end{enumerate}

The performance of multi-antenna relay for PHY-security depends
mainly on relaying protocols and schemes. For example,
amplify-and-forward (AF) and decode-and-forward (DF) are two
commonly used relaying protocols \cite{AF} \cite{DF}. AF protocol
forwards the signal polluted by noise, while DF protocol forwards
the original signal by decoding the received signal at the relay.
From the perspective of PHY-security, it is not easy to judge
which protocol is better. In general, the relaying
protocol is selected according to relaying scheme and channel
condition. In what follows, we provide an overview of various
multi-antenna secure relaying schemes.

\subsection{One-Way Relaying}

\begin{figure}[h] \centering
\includegraphics [width=0.85\textwidth] {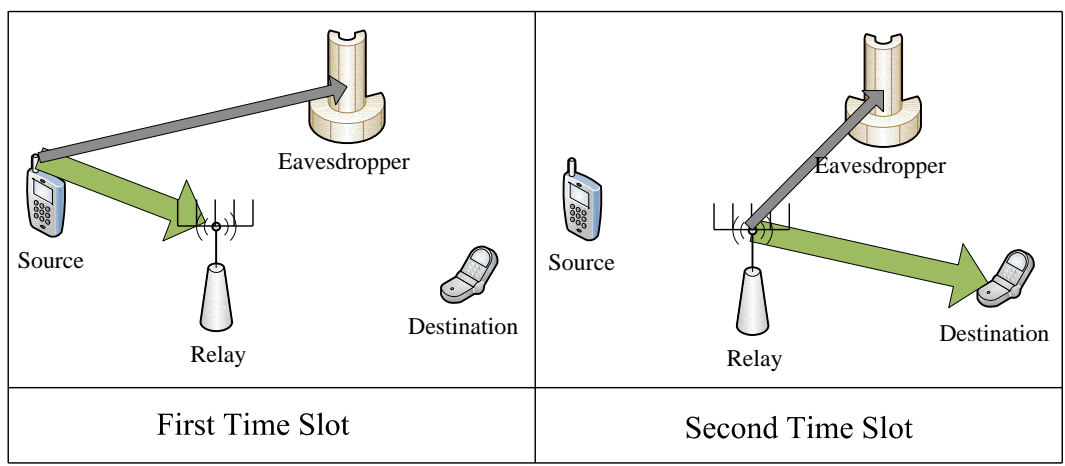}
\caption {A model of one-way secure relaying system.} \label{Fig1}
\end{figure}

One-way multi-antenna secure relaying technology is the most
popular relaying scheme. In this case, to accomplish a
transmission two time slots are required. As shown in Fig. \ref{Fig1}, during the first time slot, the source sends message to the relay, and then the relay forwards the post-processed
signal to the legitimate destination within the second time slot.
Meanwhile, the eavesdropper also receives the signals and tries to
decode them. In order to improve the secrecy performance, a
feasible way is the use of multi-antenna techniques at the relay.
For both AF and DF protocols, zero-forcing (ZF), minimum mean
square error (MMSE) or match filter (MF) receiver can be utilized
in the first time slot. Similarly, ZF, MMSE or MF transmitter is
applicable within the second time slot \cite{RelayPrecoding}.
Then, with different transceivers and relaying protocols at the
relay, there are eighteen combinations in total. According to
channel conditions, it is possible to select an optimal
combination. For example, the MMSE receiver can mitigate the
noise, and then AF is used due to its low complexity. Moreover, a
ZF transmitter can effectively decrease the
information leakage if eavesdropper CSI is available. Even without
eavesdropper CSI, the transceiver designed based only on
legitimate CSI is beneficial for secrecy performance enhancement.

Moreover, with multiple antennas at the relay, cooperative jamming
can also be used to further improve the secrecy performance.
Specifically speaking, a relay generates interference independent
of the source message (such as artificial noise) towards an
eavesdropper. In order to avoid interference to the destination,
the jamming signal is transmitted in the null space of the
relay-destination channel, making use of spatial degrees of
freedom of the multi-antenna relay. Similarly, the
source can also send the jamming signal to interfere with the
eavesdropper in the second time slot. It is worth pointing out
that there are two potential problems for cooperative jamming.
First, if CSI is imperfect, cooperative jamming may result in
residual interference to the destination. However, even with
residual interference, it may be still beneficial for wireless
security to adopt cooperative jamming as long as legitimate CSI is
sufficiently accurate. Second, the jamming signal consumes extra
power. Thus, in power-limited secure systems, it makes sense to
design an energy-efficient cooperative jamming scheme.

\subsection{Two-Way Relaying}

In a two-way relaying case, the source and the destination exchange
message with the aid of a multi-antenna relay in two time slots.
Specifically, two nodes send their signals simultaneously
to the relay during the first time slot. Then, the relay
broadcasts the post-processed mixed signal based on AF or DF
relaying protocol. Each node subtracts its transmitted signal from
the received signal, and then recovers the information from the
other node. Compared to one-way relaying, two-way relaying has two
advantages from the perspective of wireless security. First,
two-way relaying doubles the spectral efficiency of the legitimate
signal transmission. Second, the current transmission of two
signals may degrade the quality of the interception signal, since
there is no interference cancelation at the eavesdropper.

The key of two-way relaying for PHY-security lies on the design of
transceiver at the multi-antenna relay. On one hand, the
interference between two legitimate signals should be avoided,
while still guaranteeing a high spectral efficiency. To this end,
some advanced network coding techniques can be used at the relay
\cite{Twoway}. For example, physical layer network coding performs
XOR operation to the two signals on bit level after decoding the
two signals from the mixed signal, and then the desired signal can
be recovered at each source using XOR operation to the received
signal based on its own transmit signal. Moreover, ZF beamforming can
also be adopted to separate the two signals in space. On the other
hand, wireless security should be fulfilled by decreasing
information leakage to the eavesdropper. If full or partial
eavesdropper CSI is available, ZF or MMSE beamforming is an
effective way to reduce the information leakage. Otherwise, if
there is no eavesdropper CSI, cooperative jamming can be used to
enhanced wireless security.

Note that in order to decrease the complexity of
separating the mixed signal at a relay, it is likely to transform
the traditional two-slot two-way relaying to a three-slot scheme.
Specifically, one source first sends message, and then the other
source transmits its signal. Finally, the relay broadcasts the
post-processed signal. This transformed scheme may weaken the
wireless security, since there is no self-interference during the
first two time slots. Moreover, it requires a longer transmission
time. However, it may achieve a balance between security and
complexity. In addition, if the compute-and-forward
protocol is used, the relay does not need to decode each signal
from the mixed signal. Instead, it can decode a function of the
signals and forward it, which can further reduce the complexity.

\subsection{Full-Duplex Relaying}

\begin{figure}[h] \centering
\includegraphics [width=0.6\textwidth] {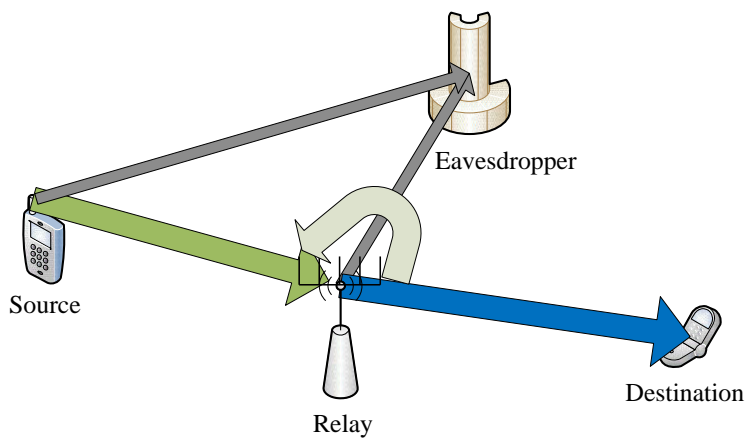}
\caption {A model of full-duplex secure relaying system.}
\label{Fig2}
\end{figure}

Both one-way and two-way relaying adopt half duplex
scheme, which separates the processes of transmitting and
receiving in time. However, if the relay can simultaneously transmit
and receive signals, namely full-duplex relaying
\cite{Full-Duplex}, as seen in Fig. \ref{Fig2}, the spectral
efficiency can be doubled with respect to one-way relaying. In
addition, the signals from the source and the relay may produce
extra interferences to the eavesdropper, and thus improve the
secrecy performance.

Although full-duplex brings great benefits for wireless security,
still it faces many challenging issues. The biggest
problem is the self-interference from the transmitted signal from
the relay to the received signal at the relay \cite{Full-Duplex}.
Due to relatively short propagation distance, the
self-interference may severely degrade the performance.
Intuitively, it is possible to cancel the interference from the
received signal, since the relay knows the transmit signal
perfectly. However, self interference is also affected by the loop
channel from the transmitter to the receiver. If the CSI for the loop channel is imperfect, the
interference cannot be cancelled completely. More importantly,
since interference has its pros and cons in PHY-security, it may not be
optimal to cancel interference completely for full-duplex
relaying. A feasible way is a joint design of transmit and
receive beams at the relay in order to achieve a fine balance between
the effects of self-interference on the legitimate and
interception signals.

\subsection{Cooperative Relaying}

If there is a strict spatial limitation at the relay, it may be
impossible to deploy multiple antennas. In this case, multiple
single-antenna relays can cooperatively assist secure
communications \cite{Cooperativerelay}. The advantages of
cooperative relaying for PHY-security are two-fold. First, these
relays are geographically distributed, then the access distance of
the destination may be shortened, and thus the secrecy performance
is improved. Second, these relays can play different roles
according to channel conditions. For example, the relays close to
the eavesdropper may act as cooperative jammers, so as to generate
strong interferences to the eavesdropper. The other relays still
forward the legitimate signal cooperatively. Compared to
cooperative jamming in a co-located multi-antenna relay, the one
with cooperative relaying has a lower complexity.

However, cooperative relaying also faces some implementation difficulties.
Specifically, cooperative relaying is in general carried out in a
distributed way. Thus, the synchronization for multiple relays is a
nontrivial task, especially for the relays with different roles.
Moreover, CSI exchange between the relays is also challenging. It
may increase overheads, and an intelligent eavesdropper can
obtain the CSI. If it succeeds, these kind of disruptive attacks can
be a serious thread and will significantly impair the secrecy
performance as a whole.

\subsection{Untrusted Relaying}

A key feature in relaying systems as described above is that they
all assumed that the relay can be trusted. In other words, the relay
will assist secure transmissions in the best way they can. However, from
recent research works, several papers have considered the use of
untrusted relays \cite{Untrusted}. In a untrusted relay model,
although the relay is a cooperative node, information intended for
the destination must be kept secret from it. Another line of works
assumed that the relay is ``malicious", i.e., the relay may try to
modify the retransmitted signal towards the destination. The use of
untrusted relay may occur in several cases. For example, in public
networks, the relays that are used for connectivity may belong to a
third party. Such relays can operate with standard protocols
although they can be unauthenticated. Malicious relay scenarios can
occur in military applications as well, where an enemy can ``pretend" to be a
cooperative node forwarding the malicious data to the destination.

The untrusted relaying has a great impact on the secrecy
performance. The achievable secrecy rate of the DF
protocol is zero, while the AF protocol can achieve a nonzero
secrecy rate. A feasible solution to the untrusted relaying is the
use of cooperative jamming. A friend sends a jamming
signal to interfere the relay, but the destination can
completely cancel the interference with aprior knowledge. Thus,
the secrecy performance in the case of the untrusted relaying can
be improved.

\section{Adaptive Resource Allocation for Multi-Antenna Secure Relaying}
In multi-antenna relay networks, there are different
types of resources, such as power, time, space and antenna
resources. These resources will affect the quality of both
legitimate and interception signals, and thus it makes sense to
allocate them according to channel conditions and system
parameters \cite{ResourceAllocation}. However, resource allocation
in secure communications is a nontrivial task. In what follows, we
discuss several key issues on resource allocation in multi-antenna
secure relaying systems.

\subsection{Adaptive Beamforming}
Beamforming has a great impact on the secrecy performance. As
aforementioned, if the legitimate signal is transmitted in the
null space of the eavesdropper channel, the eavesdropper cannot
receive any information. However, implementation of beamforming in
secure relaying systems is not easy, especially in the case
without eavesdropper CSI. In general, the source and the relay
design the beams independently. Then, the source constructs a beam
aiming at the relay if it knows legitimate CSI only. However, the
beamforming design at the relay involves multiple factors. It is quite
complex and can only be made suboptimal. On one hand, the
beamforming scheme is related to the relaying protocols. For example,
the AF protocol will forward the noise, and then it is better to
adopt a beam that may achieve a tradeoff between enhancing the
signal and mitigating the noise, i.e., ZF and MMSE. The DF
forwards the original signal, and thus MF beamforming can maximize
the signal-to-noise ratio (SNR) at the destination. On the other
hand, the receive beam in the first time slot and the transmit
beam in the second time slot should be designed jointly. Generally
speaking, the receive beam will determine the quality of the
legitimate signal, while the transmit beam can impair the
interception signal. In addition, if full-duplex relaying is
adopted, the receive and transmit beams should be designed
carefully to deal with the effect of self-interference.

\subsection{Power Allocation}
In traditional communications without security requirements, the
communication quality, e.g., transmission rate, is usually an
increasing function of transmit power. However, the power has a
side effect in secure communications. This is because increasing
the power would simultaneously improve the performance of the
legitimate and the eavesdropper channels. Thus, the power should
be allocated adaptively to the conditions of the legitimate and
the eavesdropper channels.

In secure relaying systems, power allocation becomes more
complicated, since the powers at the source and the relay are
inter-related. For example, based on the DF relaying protocol, the
legitimate channel rate is determined by the smaller of the
rates of the source-relay and the relay-destination channels.
Therefore, it does not make sense to increase the power on one side,
but fix the other. For the AF relaying protocol, increasing the
relay power may amplify the noise, resulting in performance
saturation. In addition, if a more advanced relaying scheme is
adopted, power allocation should be adjusted accordingly. As a
simple example, in full-duplex relaying systems, the relay power
directly determines self-interference, and thus the power
allocation should consider the interference cancelation scheme and
the effect of the interference on the interception signal.
Moreover, for secure relaying systems with cooperative jamming, if
the total relay power is constrained, it is necessary to
distribute the power between the forwarding signal and the jamming
signal.

\subsection{Time Allocation}
In general, the durations for the first and the second time slots
in relaying systems are equally allocated. Such an allocation
scheme is simple and asymptotically optimal, if the
relay is at the middle of the source and the destination.
However, in secure relaying systems, since the channels from the
source to the eavesdropper and that from the relay to the
eavesdropper may be quite different, equal duration allocation may
result in obvious secrecy performance loss. Specifically, if the
eavesdropper is closer to the relay, it makes sense to distribute
a longer duration to the first time slot. Intuitively, time
allocation is also related to the other system parameters, i.e.,
relaying protocol and transmit power. Hence, time allocation can
effectively enhance the secrecy performance.

\subsection{Antenna Selection}
In multi-antenna secure relaying systems, the antennas at the relay
have different effects on the secrecy performance if the channels experience
independent fadings. As mentioned in cooperative relaying, some
relays may be closer to the eavesdropper, and thus the forwarding of these
relays may lead to information leakage. Even in a co-located
multi-antenna relaying system, certain channels from a relay
antenna to the destination may experience deep fading, but the
channel to the eavesdropper may have a high gain. In this
case, the use of these antennas not only wastes the power, but also
degrades the secrecy performance.

Antenna selection in secure relaying systems is not a trivial
issue, since it is a combinatorial optimization problem from a
pure mathematical viewpoint. If the number of antennas is not so
large, it is possible to select the optimal antennas by exhaustive
searching. Otherwise, some suboptimal scheme may be used to select
the antennas. For example, if eavesdropper CSI is unavailable, the
antennas can be selected only according to the quality of the
legitimate channels.

\subsection{Relaying Protocol Switch}
There exist various relaying protocols, where AF and DF are two most
commonly used ones. In secure relaying systems, there is no
dominant protocol. As channel conditions change, the optimal
relaying protocol may also vary. Thus, it makes sense to switch the
relaying protocols according to channel conditions in order to
optimize the secrecy performance.

It is worth pointing out that the above resource allocation schemes
are interactive. For example, power allocation scheme may affect
the time allocation. Thus, it is better to optimize these
resources jointly in order to maximize the secrecy performance.

\section{Large-Scale MIMO Relaying for PLS}
In secure communications, there may be some adverse conditions,
e.g., no eavesdropper CSI and imperfect legitimate CSI. In this
context, if the interception distance is relatively short, then even with a
multi-antenna relay, the secrecy performance may be very poor. As
a result, it is difficult to provide secure, reliable and QoS
guaranteed communications.

To solve the problem with short-distance interception in secure
communications, we recently proposed to use large-scale MIMO
(LS-MIMO) relaying technology to enhance wireless
security significantly \cite{LS-MIMOrelay}. LS-MIMO can generate a very
high-resolution spatial beam, making use of a large number of
antennas. Thus, on one hand, the performance of the legitimate
channel can be improved enormously due to the high array gain. On
the other hand, the information leakage to unintended users can be made
very small. Especially, as the number of antennas tends to be infinity,
the information leakage is negligible. Then, even under adverse
conditions, it is still likely to achieve a good secrecy
performance. Additionally, compared to traditional multi-antenna
secure relaying technologies, LS-MIMO secure relaying technology
offers several appealing advantages. First, LS-MIMO simplifies the
signal processing, and even with a low-complexity transceiver at the
relay, i.e., maximum ratio combination (MRC) and maximum ratio
transmission (MRT), it is still able to achieve a good performance.
Second, it is easy to improve the secrecy performance by adding
antennas only at the relay. Third, due to channel hardening in
LS-MIMO systems, the performance analysis and optimization becomes
simpler. In what follows, we show the performance gain of several
adaptive resource allocation schemes in LS-MIMO secure relaying
systems through numerical simulations.

Let us consider a one-way secure relaying system, where the source communicates with the destination
with the aid of an LS-MIMO relay. The number of antennas $N_R$ at
the relay is very large, e.g., $N_R=100$ or even bigger. The relay
has full CSI about the source-relay channel through channel
estimation, imperfect CSI of the relay-destination channel due to channel reciprocity, but no CSI of the
relay-eavesdropper channel. The eavesdropper is closer to the relay, but
not the source, since it assumes that the signal is from the
relay directly. We use $\alpha_{S,R}$, $\alpha_{R,D}$, and $\alpha_{R,E}$ to
denote the normalized path loss of the source-relay channel, the
relay-destination channel, and the relay-eavesdropper channel,
respectively. Note that we take secrecy outage capacity as the
performance metric, since eavesdropper CSI is unavailable. Secrecy
outage capacity is defined as the maximum transmission rate, while
secrecy outage probability needs to satisfy a given constraint. In this
manuscript, the bound on secrecy outage probability is set to 0.05.

\begin{figure}[h] \centering
\includegraphics [width=0.865\textwidth] {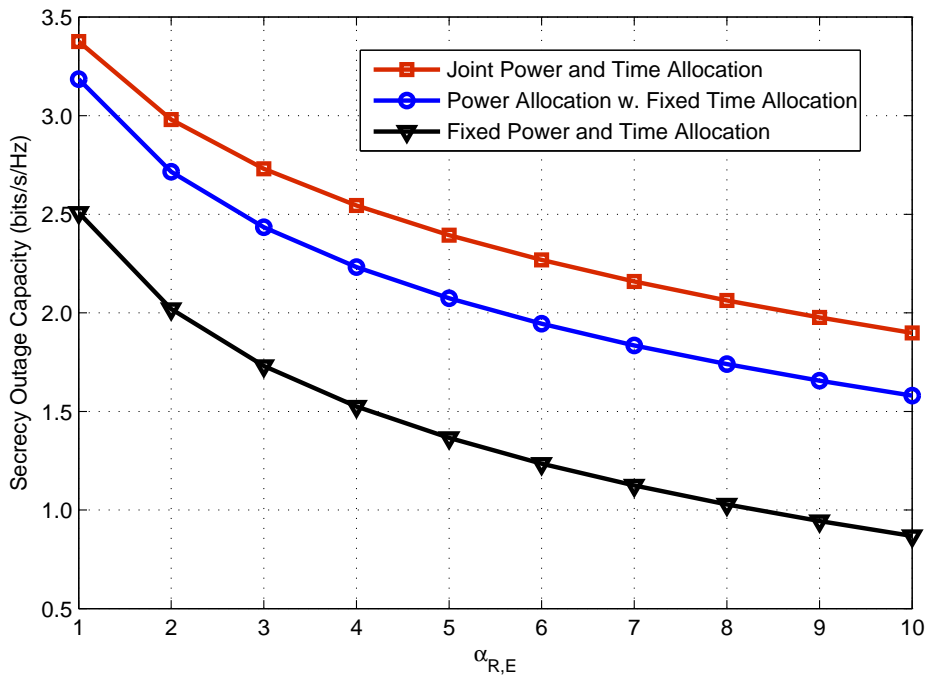}
\caption {Secrecy performance comparison with joint and fixed power allocation.} \label{Fig3}
\end{figure}

First, we show the performance gain of joint resource
allocation over fixed resource allocation scheme in a DF
LS-MIMO secure relaying system. For analysis convenience, we normalize
$\alpha_{S,R}=\alpha_{R,D}=1$, and use $\alpha_{R,E}>>1$ to
represent short-distance interception. We consider the
optimization of source power, relay power and duration ratio
between the first and the second hops. As seen in Fig. \ref{Fig3},
joint power and time allocation scheme obviously performs better
than power allocation with fixed time allocation scheme. This is
because duration ratio between the two hops also has a great
impact on the secrecy performance. For example, if the
eavesdropper is close to the relay, it is better to use a small
duration in the second hop. Meanwhile, transmit powers at the
source and the relay also affect the duration ratio. Thus, it
makes sense to optimize power and time jointly. Moreover, if both
power and time are fixed regardless of channel conditions and
system parameters, there will be more performance loss. Thus,
joint resource allocation can effectively improve the secrecy
performance.

\begin{figure}[h] \centering
\includegraphics [width=0.865\textwidth] {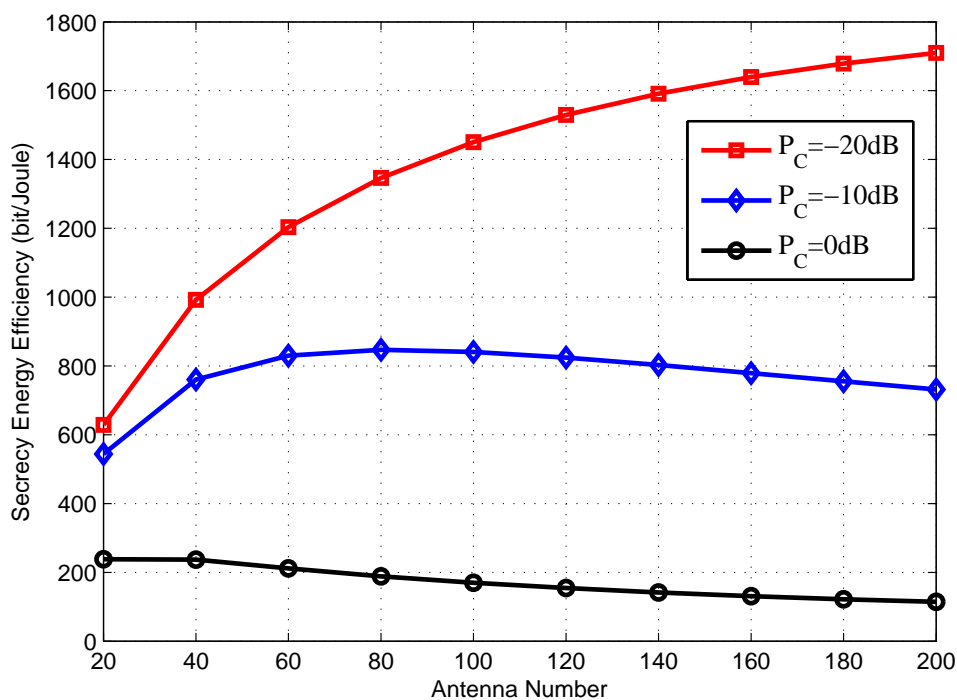}
\caption {Secrecy energy efficiency of a secure relaying system with
different numbers of antennas.} \label{Fig5}
\end{figure}

Then, we examine the impact of the number of antennas at the relay
on the secrecy performance in an AF LS-MIMO secure relaying system
with $\alpha_{S,R}=\alpha_{R,D}=\alpha_{R,E}=1$. Intuitively,
adding more antennas can always improve secrecy outage
capacity, but also increases resource consumption, such as
power. In this case, we take secrecy energy efficiency as the
performance metric, which is defined as the ratio of secrecy
outage capacity and total consumed power, including transmit
power, circuitry power per antenna, and basic power independent of
the number of antennas. In Fig. \ref{Fig5}, we use $P_C$ to denote
the circuitry power per antenna. It is found that if $P_C$ is very
small, i.e., $P_C=-20$ dB, adding more antennas is always helpful to
increase the secrecy energy efficiency. However, with $P_C=-10$
dB, the energy efficiency first increases and then decreases as
the number of antennas increases. This is because when the number
of antennas is small, adding more antennas can increase the
secrecy outage capacity significantly. However, when the number of
antennas is relatively large, although adding more antennas can further
increase the secrecy outage capacity, the consumed power increases
sharply. Thus, it makes sense to select the optimal number of
antennas in order to maximize the energy efficiency.

In summary, adaptive resource allocation can effectively improve the
secrecy performance. LS-MIMO secure relaying technology simplifies
the signal processing and thus it is possible to optimize the utilization of
different resources jointly, such as power and time. Therefore, wireless
security can be enhanced significantly.

\section{Future Research Directions}
Wireless security is always a critical issue. Although the
introduction of multi-antenna relay can improve the
secrecy performance effectively, there are many challenges remained to be tackled. As our
future works, we intend to solve these problems in the
following directions to enhance wireless security further.

\subsection{Mobile Relay}
The position of the relay has a great impact on the performance,
especially in secure mobile communications. As channel conditions
change, the optimal position of the relay may also need to vary
accordingly. Hence, a fixed relay may result in an obvious
performance loss. If it is a vehicular relay, it
should be able to flexibly move the position and select the
secrecy scheme. For example, the relay moves closer to the
eavesdropper to strengthen the interference to the
eavesdropper through cooperative jamming. However, it is not a
trivial task to design the scheme of mobile relay. First, it
requires full CSI, which increases the overhead. Second, there is
a balance between secrecy performance and implementing complexity,
which is again an open issue.

\subsection{Multiuser Access}
In modern communications, multiuser concurrent transmission is commonly
used to improve the spectral efficiency. For example, the LTE system
supports multiple users' access through a relay. In multiuser secure
relaying systems, multiuser transmission faces several challenges. On one hand, the inter-user interference degrades the secrecy performance. On the other hand, the inter-user interference
can be used to impair the interception signal. Thus, it is necessary
to design effective user scheduling and precoding schemes to
optimize the secrecy performance.

\subsection{Combination of Encryption and PHY-Security}
PHY-security emphasizes mainly on pure signal processing
techniques, while high-layer cryptographic techniques works well
independent of channel conditions. In secure relaying systems,
the CSI may be imperfect or even unavailable, and then we can
integrate cryptographic techniques into the transceiver design.
Combining cryptographic techniques and PHY-security offers another
way to improve the secrecy performance significantly.

\section{Conclusion}
This article provides an overview of multi-antenna relaying
technologies in PHY-security, and discusses the opportunities and
challenges in the design of secure relaying systems. Through
analyzing the characteristics of secure relaying communications,
we give a comprehensive tutorial on adaptive resource allocation
schemes to further improve the secrecy performance. To solve the
problem with short-distance interception under adverse conditions,
we proposed to use LS-MIMO relaying technology and showed its
effectiveness through simulations. Finally, we identified several
research directions as our future works.

\end{spacing}
\end{document}